# Measuring the Capacitance of Individual Semiconductor Nanowires for Carrier Mobility Assessment

Ryan Tu[3], Li Zhang[1], Yoshio Nishi[2], and Hongjie Dai[1*]

[1]Department of Chemistry and Laboratory for Advanced Materials, Stanford University, Stanford, CA 94305, USA

[2]Department of Electrical Engineering, Stanford University, Stanford, CA 94305, USA

[3]Department of Materials Science and Engineering, Stanford University, Stanford, CA 94305, USA

**Abstract**

Capacitance-voltage characteristics of individual germanium nanowire field effect transistors were directly measured and used to assess carrier mobility in nanowires for the first time; thereby removing uncertainties in calculated mobility due to device geometries, surface and interface states and gate dielectric constants and thicknesses. Direct experimental evidence showed that surround-gated nanowire transistors exhibit higher capacitance and better electrostatic gate control than top-gated devices, and are the most promising structure for future high performance nanoelectronics.

[*] Email: hdai@stanford.edu

Much excitement has been generated in recent years about semiconductor nanowires (NWs) for future high performance electronics.[1-3] The potential of higher carrier mobility in NWs than in bulk materials due to 1D transport are intriguing.[4] Various NWs with mobility near or higher than bulk values have been reported, including Si, Ge, and InAs.[1, 5, 6] However, a main issue concerning this topic has been the lack of direct capacitance measurements in NW field effect transistors (FETs), which precludes confirmation of the carrier mobility values as they scale as the inverse of gate capacitances. Thus far, mobility analysis for NWs has relied on capacitances derived from modeling or simulation, which could be inaccurate due to uncertainties in interface states, defects and experimental variations causing fluctuations in the property and geometry of dielectric materials. Capacitance-voltage (C-V) measurements of planar metal-oxide-semiconductor (MOS) stacks have been widely used to investigate the properties of dielectric layers and interfaces with various semiconductors.[7] Similar measurements would be invaluable to semiconductor NWs but have remained illusive thus far due to difficulties in measuring ultra-small capacitances intrinsic to nanowires (aF to fF level) over large background parasitic capacitances (pF level).

Based on novel measurement and device designs, we perform the first direct measurements of gate capacitances in nanowire FETs in both top-gate and surround-gate geometries for various NW channel lengths. This enables evaluation of hole mobility in Ge NWs using experimentally determined capacitance data and sheds light into the validity of simulation. The results are important to elucidating the intrinsic electrical properties of semiconductor nanowires in general with little ambiguity and devising optimal performance NW electronics.

Top-gated (Figure 1a, with Si back-gate controlling two ~1.5 μm long NW segments underlapping[8] the top-gate near the source and drain contacts) and self-aligned surround-gated (Figure 1b, underlapped region ~ 40nm)[3] individual Ge NW FETs with ~1.25nm amorphous $SiO_x$ passivation layer[9] and ~4.55nm $Al_2O_3$ high-κ dielectrics (Figure 1c) were fabricated on $SiO_2$/Si substrates with gate lengths $L_g$~0.7μm-4μm. The diameters of the nanowires were ~20nm ± 3nm and were measured by atomic force microscopy for each device used for capacitance measurements. The current-gate voltage ($I_d$-$V_g$) transfer characteristics of the devices exhibited depletion-mode p-type FET behavior (Figure 2a & 3a) with on/off ratio ~$10^5$ at various drain biases ($V_d$) and subthreshold slope S~110-130mV/decade.

For measuring small gate capacitance values (~0.1-1fF) of NW FETs, we devised an approach[3] based on a method recently developed for carbon nanotubes[10] to reach a capacitance sensitivity limit ~30aF (Figure 1d). Capacitance measurements were performed with a capacitance bridge (Andeen-Hagerling, Model 2550A, 100mV signal) in a variable temperature cryoprobe station (Lakeshore, Model FWP6). The capacitance bridge cancels any stray capacitances between the back-gate (grounded) and the S,D, and G electrodes. Direct capacitances between the G and S/D electrodes are screened by the close proximity (200nm) of the back-gate. Further details of the bridge operation are discussed in a similar method for measuring quantum capacitance in carbon nanotubes.[10] During measurements, a grounded copper plate was positioned with a micromanipulator between the S/D and G pads (Figure 1d) to shield inter-probe capacitances. We found that this step was key to reduce the background capacitance from ~10fF down to the sensitivity limit of the capacitance bridge (~30aF) and was essential in eliminating the remaining parasitic capacitances that were not cancelled out by the bridge circuit. We verified a stable background capacitance by measuring dummy FET structures without the presence of any nanowires.

For top-gated GeNW FETs ($L_g$~3µm), the underlapped segments of the NWs near the S and D were switched from electrically-on to -off (by varying the Si back-gate $V_{bg}$ from -5V to 5V) while the top-gated channel was turned-on by a fixed $V_g$=-1V, during which the measured capacitance between S/D and G evolved (Figure 2b) from ~ 1.02fF to a negligible background capacitance of ~30aF. At negative $V_{bg}$, the underlapped regions of the device behave like low resistance contacts (electrically on) whereas at positive $V_{bg}$, the resistance increases until the bridge does not have enough time to charge the capacitor and the measured capacitance falls to the background level. This condition is satisfied when 1/RC << the small signal frequency of the bridge (1 KHz). The ~1.02fF provided a direct measurement of the gate capacitance between the S/D and G for the on-state of the top-gated NW FET channel (at $V_g$ = -1V). We found that this condition was not satisfied unless device were cooled to 200K, where the overall resistance increases due to the presence of Schottky S/D contacts. At low temperatures, the off-state resistance of the FETs is sufficiently high for the measured capacitance to fall to the background value. Measurements under various top-gate voltage with the underlapped segments fully turned-on by $V_{bg}$=-5V led to gate capacitance vs. top-gate ($C_g$-$V_g$) curve (Figure 2c), representing the first C-V data for semiconductor nanowires. A lower temperature (T=150K) is required such that the top gate can sweep the device from on- to off-state with near zero capacitance. The on-state capacitance decreases slightly due to a limit in carrier density through the Schottky contacts at such low temperatures. Capacitance bridges operating at higher frequencies are desired and will allow for high frequency C-V measurements of NW devices at a wider range of temperatures. This will then provide C-V data for nanowires analogous to those of planar MOS structures. We further measured the gate capacitance as a function of $L_g$ from 0.7µm to 4µm (Figure 2d) and found a capacitance per unit length of ~0.41fF/µm. Interestingly, extrapolation of the $C_g$-$L_g$ curve to $L_g$=0µm gave a non-zero capacitance of ~310aF. This was attributed to the fringe capacitance that originated from coupling between the metal gate electrode to the underlapped segments of the GeNW on both sides of the gate line.

Next, we carried out electrical and capacitance measurements for surround-gate GeNW FETs[3] (in Figure 1b) fabricated using NWs with the same diameter, surface passivation, and $Al_2O_3$ gate dielectric layer as for the top-gated FETs (in Figure 1a). These devices exhibit on/off ~$10^5$, S~120mV/decade and on-state current of ~4µA at $V_d$=-1.5V for $L_g$=4µm channel devices (Fig.3A). For surround-gate GeNW FETs; however, we were unable to find a suitable temperature range to measure a complete $C_g$-$V_g$ curve due to lack of significant underlapped regions, which were found necessary to shut off the capacitance measurement. Though varying the temperature allows some tuning of the contact resistance, it was not sensitive enough to find a single temperature whereby the capacitor could be turned on and off. Instead, $C_g$ is recorded as the on-state room temperature ($V_g$=-1V) capacitance measurement. We confirmed that the background capacitance in the off-state ($V_g$=1V) at low temperature was unchanged at ~30aF, which suggests no noticeable capacitance contribution from the close proximity of the surround-gate to the S/D contacts. $C_g$-$L_g$ measurements (Figure 3b) revealed that the capacitance per unit gate-length for surround-gate devices was ~27% higher than that of top-gated GeNW FETs (0.52fF/µm vs. 0.41fF/µm), providing an experimental proof of better electrostatic control of nanowires with surrounded-gate.[4] Extrapolation of the $C_g$-$L_g$ curve to $L_g$=0µm gave a near zero fringe capacitance, consistent with the very short underlapped segments (~40nm vs. ~1.5µm in the top-gated devices) in the surround-gate FETs due to self-aligned S/D and G.[3]

We extracted carrier mobility in surround-gated GeNWs by using the gate-capacitances obtained through direct experimental measurements. The mobility was calculated in the linear region by,[7]

$$\mu = \frac{dI_d}{dV_g} \frac{1}{C_g V_d} L_g^2 \qquad (1)$$

We found an low-field hole mobility of 400cm$^2$/Vs at $V_d$=-10mV for the $L_g$=4µm surround-gated GeNW FET (Figure 3). The extracted mobility was accurate based on experimentally determined capacitance, but should be taken as a lower bound of the hole mobility in our surround-gate Ge NWs since the contacts were non-ohmic with finite Schottky barriers (evidenced by reduced thermal emission conduction at lower temperatures)[3] and that the short but finite gate underlapped NW segments introduced serial resistance into the FETs. The serial resistance was high for top-gated FETs with long underlapped regions, and was corrected for in our carrier mobility analysis.[3]

Our extracted low-field mobility of ~400cm$^2$/Vs was within the range of reported hole mobility of 100-770cm$^2$/Vs in planar Ge FETs.[11-14] It was also within the wide range of reported hole mobility values for GeNW FETs (20-730cm$^2$/Vs) derived from indirect capacitance estimates based on modeling or simulation.[1, 15, 16] Our result was credible with no uncertainty in gate-capacitance. The 400cm$^2$/Vs hole-mobility in Ge NWs was ~2 times higher than that of Si devices, confirming the potential of GeNWs for high performance p-type FETs. However, the result was significantly below the hole mobility of ~1900cm$^2$/Vs in bulk Ge and that of ~770 cm$^2$/Vs reported for Ge-crystalline Si core-shell NWs[1], likely due to the non-optimal amorphous SiO$_x$ surface passivation used in our devices.

Our experimental results can shed light into capacitance modeling widely used for nanowires. Transmission electron microscopy (TEM) was employed to accurately determine the thickness of the Al$_2$O$_3$ and the SiO$_x$ interfacial layer (Figure 1c). We determined the dielectric constant of ALD deposited Al$_2$O$_3$ to be κ~7.3 by fabricating planar MOS stacks on Si and performing standard C-V measurements.[3] C-V measurements were also performed on planar MOS stacks on p-type Ge wafers with the same Al$_2$O$_3$ dielectric and SiO$_x$ passivation layers[3] as in the NW FETs. We found that the SiO$_x$/Ge interface and SiO$_x$ layer gave an effective dielectric constant of κ~1.7 for the ~1.25nm SiO$_x$ layer, significantly lower than the dielectric constant of κ~3.9 for SiO$_2$ likely resulted from a combination of SiO$_x$/Ge interface states and SiO$_x$ porosity.[17] With these detailed characteristics, we found that 2-D electrostatic finite element simulations of top-gated GeNWs were able to reproduce experimentally measured capacitances provided void spaces were considered under the NW and gate metal (Figure 4a & 4b). Existence of such voids was reasonable and caused by the NW on substrate masking the incidence of metal-atom flux during the metal gate sputter-deposition process.[18]

Comparison between experiment and simulation identified a ~10° angle of incidence for the gate-metal deposition flux (Figure 4b left panel). Such an effect would be difficult to quantify by other means for inclusion into capacitance modeling (a ~26% error in capacitance would be introduced without considering the void). This highlighted the importance of direct capacitance measurement for NW devices.

For surround-gated NW FETs, simulations were able to reproduce the experimentally determined gate capacitances (Figure 4a). Importantly, this agreement came as a result of using carefully characterized geometrical and dielectric and interfacial parameters obtained by microscopy and C-V measurements on planar MOS stacks. Thus, without the capability of direct capacitance measurements for NW devices (though always preferred), one should characterize corresponding planar MOS structures and use the results to model the capacitances of NW devices with the least uncertainty in dielectric and interfacial structures and properties.

In summary, we have performed the first direct capacitance measurements for assessing carrier mobility in nanowires. Experiments confirm that surround-gate structure affords higher gate capacitance and optimal electrostatic control of NWs and presents the most promising approach to high performance NW electronics.

**Acknowledgements:**
We thank Dr. Shahal Ilani and Prof. P. McEuen for helpful discussions regarding capacitance measurements. We also thank Andeen-Hagerling for use of capacitance bridge and Agilent Technologies for use of a B1500 for electrical measurements. This

9work was supported by MARCO MSD, Intel, Stanford INMP and an NSF Graduate Research Fellowship (R.T.).**Supporting Information Available:**

This material is available free of charge via the Internet at http://pubs.acs.org

**Figure Captions**

**Figure 1** Gate capacitance measurement for individual NW FETs. (a) & (b) Schematics and scanning electron microscopy (SEM) images of top-gated (underlapped NW length on S and D side ~1.5µm) and surround-gated GeNW FET respectively. Scale bars: 2µm. The surround-gate electrode spans nearly the entire distance between S/D electrodes with a small gap (~40nm) on each side formed by a self-aligned process. (c) A TEM image of the dielectric layers deposited on a GeNW. Combined thickness of the $SiO_x$ + $Al_2O_3$ layers is 5.8nm with a standard deviation of 1.6Å. (d) Schematic of capacitance measurement setup. A grounded copper plate is positioned between the S/D and G probe tips, which reduces background capacitance from ~10fF to ~30aF. In practice, the copper plate extends higher than the probe tips for effective shielding. A top-gate or surround-gate voltage ($V_g$) and a back-gate voltage ($V_{bg}$) can be applied to the backside of the substrate during measurement.

**Figure 2** Electrical and $C_g$-$V_g$ data of a top-gated GeNW FET ($L_g$=3µm). (a) Room temperature transfer characteristics $I_d$-$V_g$. Inset: $I_d$-$V_d$ curves at $V_g$ = -1, -0.5, and 0V. The current increase at high $V_g$ can be attributed to band-to-band tunneling. (b) Room temperature and low temperature capacitance data (left axis) during switching the back-gate $V_{bg}$ from -5V to 5V (right axis) to turn the underlapped NW segments at the S/D from electrically-on to -off state. $V_{bg}$ is varied as a step function from -5V to 5V while the top-gate $V_g$ is fixed at -1V to maintain an on-state for the transistor channel under the top-gate. On-state capacitances at room and low temperature were identical. At room temperature, the channel off-state capacitance does not approach background capacitance

(near zero) since the off-state electrical resistance (~10MΩ) is not sufficiently high to shut off charge to the channel at the operating frequency of the bridge of 1kHz. To reach the off-state, the channel resistance ($R_c$) must be high such that $1/R_c C_g \ll$ 1kHz. At low temperature, the conductance of both on- and off-state decreases due to a finite Schottky barrier at the contacts, and the measured off-state capacitance approaches zero with a high off-resistance of ~10TΩ. (c) Capacitance-voltage ($C_g$-$V_g$) curve of the device at T=150K. $V_{bg}$ is fixed at -5V while top-gate $V_g$ is varied from -1V to 1V. (d) $C_g$ vs. channel length $L_g$ from 0.7µm to 4µm. The error bars represent the range of capacitance measurements of multiple devices of a particular gate length.

**Figure 3** Electrical and capacitance measurements of a surround-gate GeNW FET ($L_g$~4µm). (a) Room temperature transfer characteristics $I_d$-$V_g$. Inset: $I_d$-$V_d$ curve with $V_g$ = -1, -0.5, and 0V respectively. (b) $C_g$ vs. $L_g$ for surround gate FETs. The error bars represent the range of capacitance measurements of multiple devices of a particular gate length. Slope of the linear fit line (0.52fF/um) represents the average capacitance per unit length of the surround gate GeNW FET. The intercept near zero confirms minimal underlap due to the self-aligned nature of the fabrication process. $C_g$-$V_g$ curves could not be measured due to lack of significant underlapped regions, which were found necessary to shut off the capacitance measurement [3]. Instead, $C_g$ is recorded as the on-state room temperature ($V_g$=-1V) capacitance measurement. Measurements of the off-state ($V_g$=1V) at low temperature confirm that the background capacitance is ~30aF and negligible compared to $C_g$.

**Figure 4** Experimental and simulations of capacitances of top-gated and surround-gated GeNW FETs. (a) Comparison of experimental and simulation data for top-gated and surround-gated GeNW FETs with various gate lengths. The experimental data for top-gated GeNW FET are represented by circles and shifted downward by the fringe capacitance amount for comparison purposes. The simulation data for top-gated devices have considered void spaces between the NW and the substrate as shown in (b) left panel, caused by NW shadowing the gate-metal deposition during the device fabrication process. (b) Equipotential lines from 2-D finite element simulations of top-gated and surround-gated GeNW FETs. Equipotential lines are shown with red representing highest potential and purple representing lowest potential. The capacitance simulation (Field Precision Software EStat 6.0) employs finite element methods on triangular meshes of a 2D cross section of the nanowire device. Capacitance (c) is calculated by analyzing the total energy (E) in the dielectric surrounding the nanowire and using the equation $E=1/2CV^2$, where V=1V is the simulated potential between NW and gate.

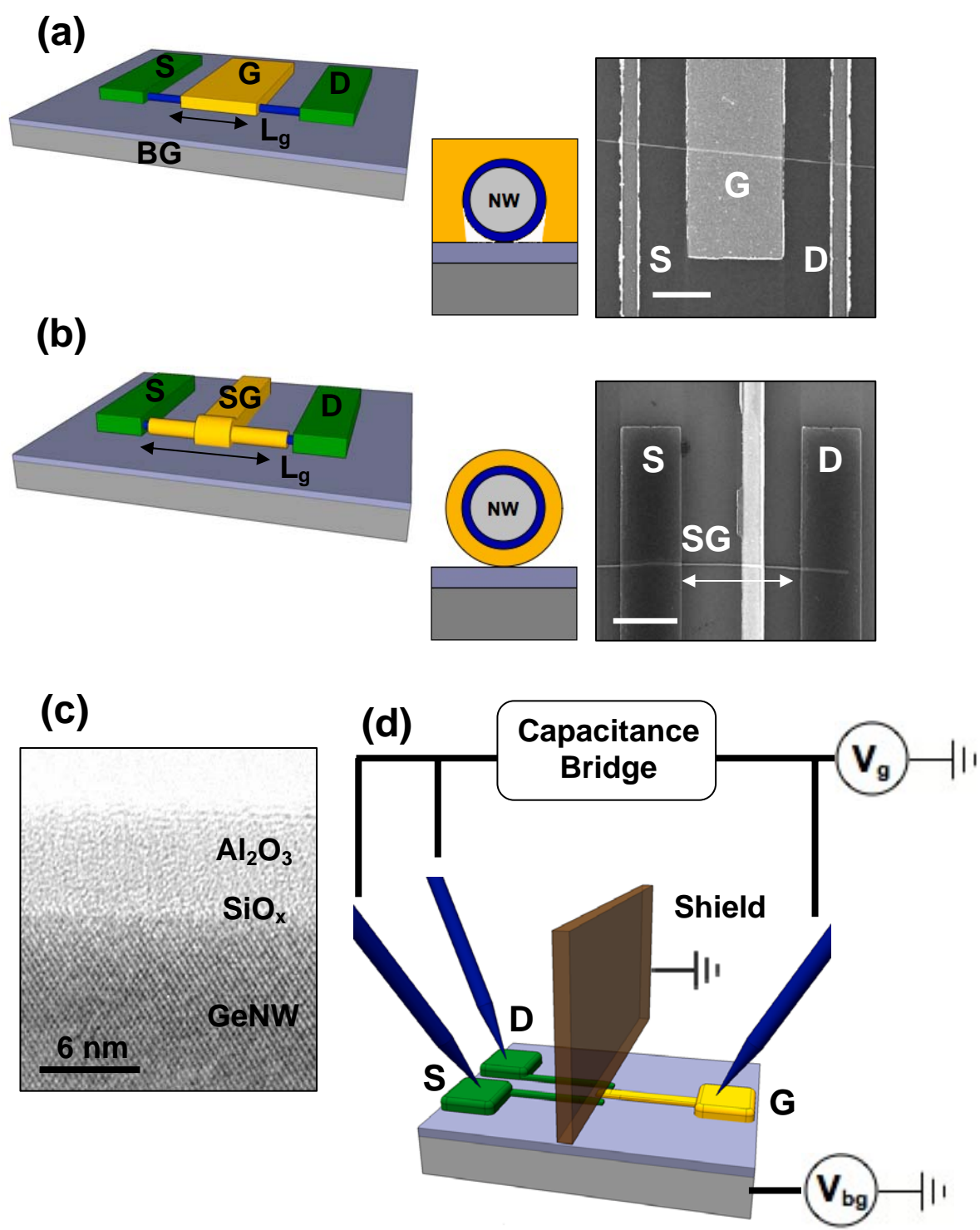

**Figure 1**

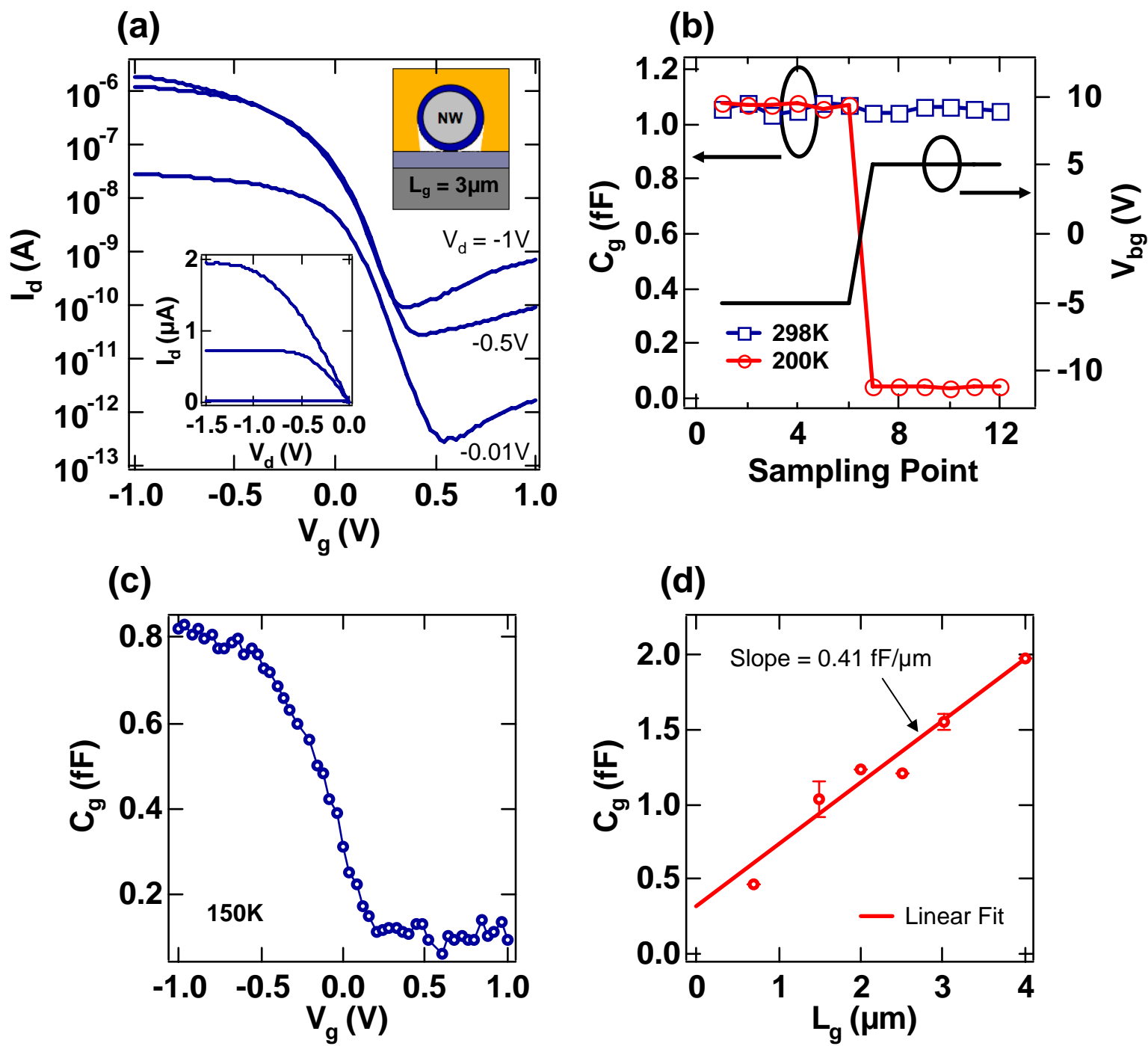

**Figure 2**

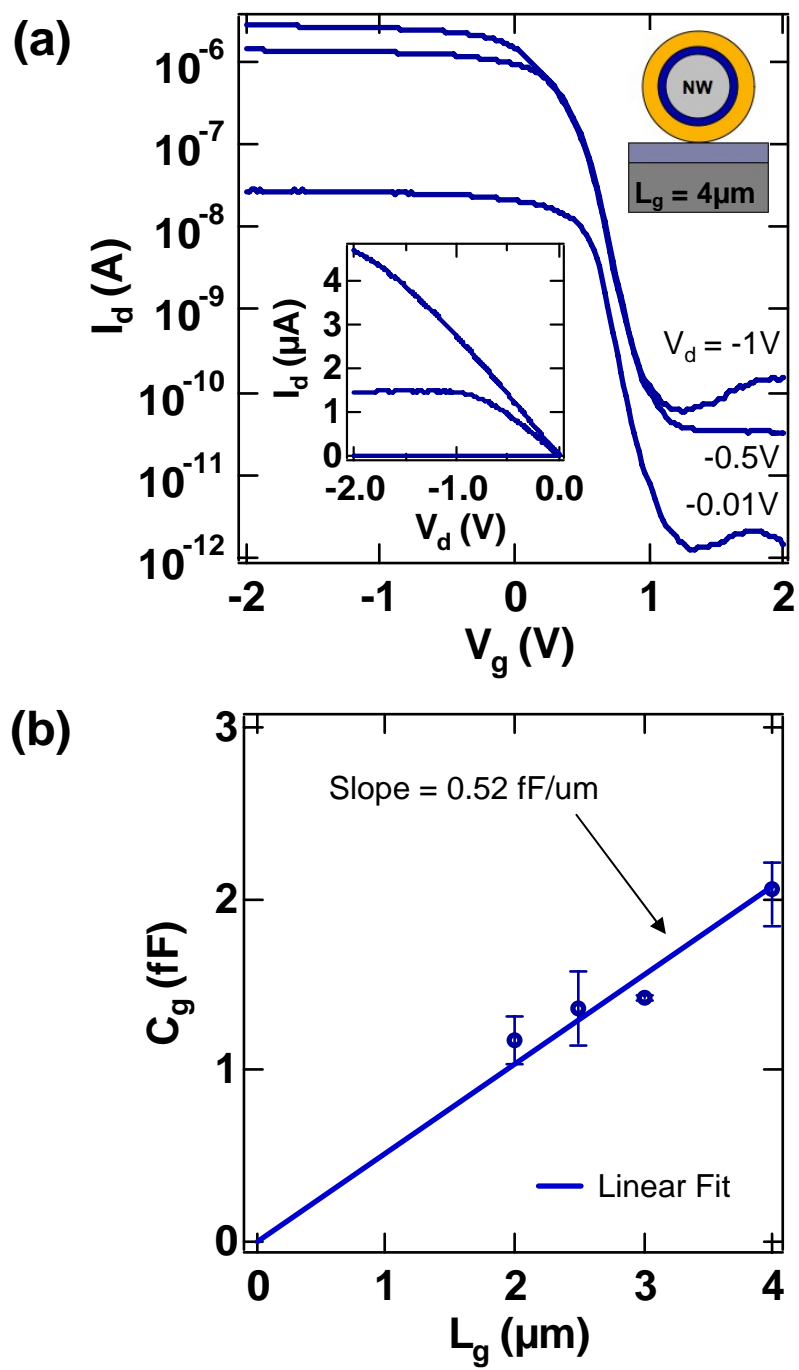

**Figure 3**

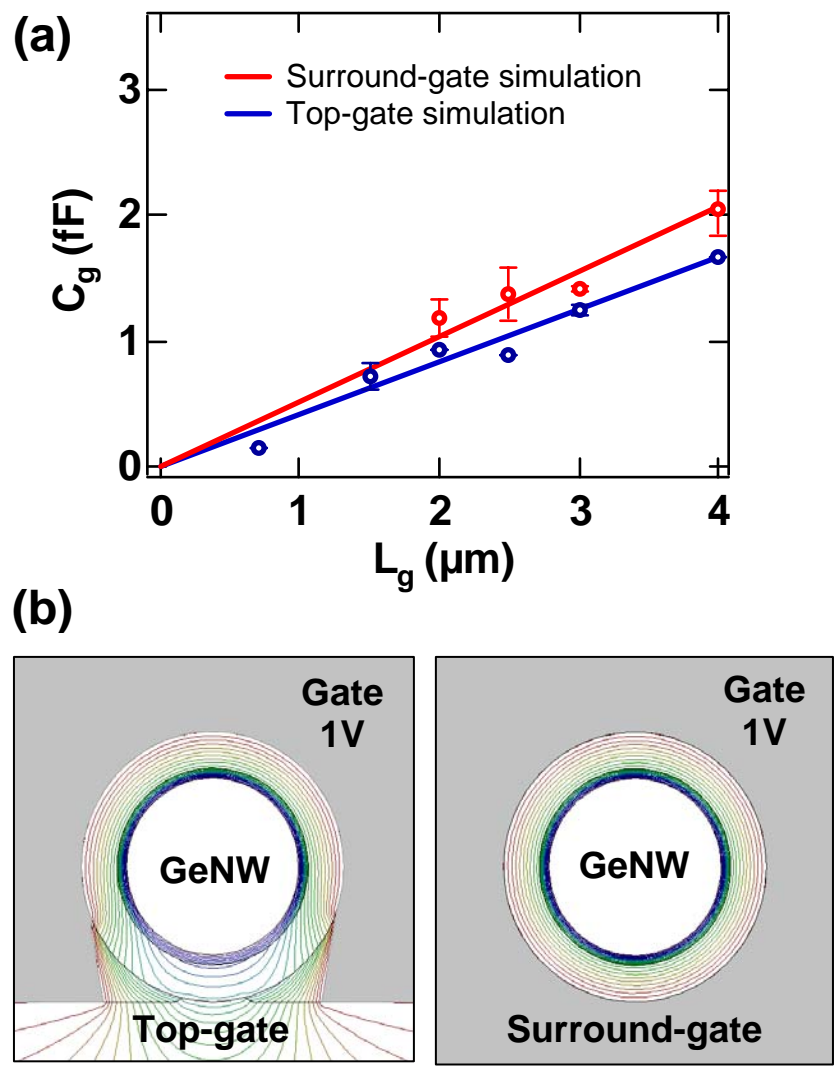

**Figure 4**